\input{aipcheck}

\documentclass[
    ,final            % use final for the camera ready runs
  ]
  {aipproc}

\layoutstyle{6x9}

\newcommand{\beq}{\begin{equation}}
\newcommand{\eeq}[1]{\label{#1} \end{equation}}

\usepackage{amsmath}
\usepackage{amsfonts}
\usepackage{amssymb}
\usepackage{graphicx}
\usepackage{epsfig}

\begin{document}

\title{Multi-Module Modeling of Heavy Ion Reactions\\
        and the 3rd Flow Component$^\dagger$}

\author{\underline{L.P. Csernai}$^{1,2}$
A. Anderlik$^{1}$,
Cs. Anderlik$^{1}$,
V.K. Magas$^{3}$,\\
E. Moln\'ar$^{1}$,  
A. Nyiri$^{1}$,    
D. R\"ohrich$^{1}$,  and    
K. Tamosiunas$^{1}$
}{
  address={
$^1$ {\it Department of Physics and Technology}\\
{\it University of Bergen, Allegaten 55, 5007 Bergen, Norway}\\[1.2ex]
$^2$ {\it KFKI Research Institute for Particle and Nuclear Physics}\\  
{\it P.O.Box 49, 1525 Budapest, Hungary}\\[1.2ex]
$^3$ {\it Departamento de F{\'{\i}}sica Te{\'o}rica and IFIC, Centro Mixto,}\\ 
{\it Institutos de Investigaci{\'o}n de Paterna - Universidad de Valencia-CSIC}\\ 
{\it Apdo. correos 22085, 46071, Valencia, Spain}\\[1.2ex]
$^\dagger$ {\it Invited talk, presented at the "Hadron-RAMP 2004" meeting,}\\
{\it March 28 to April 3, 2004, Angra dos Reis, (Rio de Janeiro), Brasil}  
}}

%\author{<author2>}{
%  address={<common address for author2 and author3>}
%}

%\author{<author3>}{
%  address={<common address for author2 and author3>}
%}

\begin{abstract}
 
Fluid dynamical calculations with QGP 
showed a softening of the directed flow while with hadronic matter this
effect is absent. On the other hand, we indicated that a third flow
component shows up in the reaction plane as an enhanced emission, which 
is orthogonal to the directed flow. This is not shadowed by the 
deflected projectile and target, and shows up at measurable rapidities,
$y_cm = 1-2$. To study the formation of this effect initial stages of 
relativistic heavy ion collisions are studied. An effective string 
rope model is presented for heavy ion collisions at RHIC energies. 
Our model takes into account baryon recoil for both target and projectile, 
arising from the acceleration of partons in an effective field. The typical
field strength (string tension) for RHIC energies is about 5-12 GeV/fm, 
what allows us to talk about "string ropes". The results show that  
QGP forms a tilted disk, such that the direction of the largest pressure 
gradient stays in the reaction plane, but deviates from both the beam 
and the usual transverse flow directions. The produced initial state 
can be used as an initial condition for further hydrodynamical calculations. 
Such initial conditions lead to the creation of third flow component.
Recent $v_1$ measurements are promising that this effect can be used
as a diagnostic tool of the QGP. Collective flow is sensitive to the early 
stages of system evolution. To study the sensitivity of the flow signal, 
we have calculated flow harmonics from a Blast Wave model, a tilted, 
ellipsoidally expanding source. We studied recent experimental techniques 
used for calculation of the $v_n$ Fourier coefficients and pointed out 
a few possible problems connected to these techniques, which may impair the
sensitivity of flow analysis.

\end{abstract}

\maketitle

%\tableofcontents

\section{Introduction}\label{intro}

In high energy heavy ion reactions many thousand particles are
produced in the reaction, so, we have all reasons to assume that in a
good part of the reaction the conditions of the local equilibrium and 
continuum like behavior are satisfied. The initial and 
final stages are, on the other hand, obviously not in
statistically equilibrated states, and must be described separately,
in other theoretical approaches. The different approaches, as they
describe different space-time (ST) domains and the corresponding
approaches can be matched to each other across ST hyper-surfaces or
across some transitional layers or fronts. The choice of realistic
models at each stage of the collision, as well as the correct coupling of
the different stages or calculational modules are vital for a
reliable reaction model.

The fluid dynamical (FD) model plays a special role among the 
numerous reaction models, because it can be applied only 
if the matter reaches local thermal equilibrium. If this
happens the matter can be characterized by an Equation of
State (EoS). This is what we are actually looking for in these
experiments and this is why fluid dynamics is so special.

\section{Measurement of Colective Flow}

In relativistic heavy ion 
collisions collective flow was predicted to occur  
already in the early 1970s
\cite{SMG,CJTW}
and proved to exist beyond doubt in 1984 first by the Plastic
Ball collaboration at LBL \cite{PB84}. 
Then, using the method worked out by 
Danielewicz and Ody\-ni\-ecz based on the physical properties of
heavy ion collisions \cite{DO85}, 
the directed collective flow was possible to detect in smaller samples
and in a wide range of detectors.

A variety of collective flow patterns were detected, 
the ``squeeze out'', the ``elliptic flow'', the ``anti flow'' or ``3rd flow component'',
etc. Anisotropic flow is defined as azimuthal asymmetry in particle distributions with 
respect to the reaction plane, spanned by the beam direction and the impact parameter vector. 
The increasing complexity of flow patterns naturally led to 
attempts to classify the flow patterns in a more systematic way, on principal
ground, and not just following the new experimental observations.

A formal classification of the azimuthal asymmetries was recently introduced
via the coefficients, $v_n$, of the Fourier expansion of the azimuthal 
distribution of particles
\cite{PV98}:
\begin{equation}
E \frac{d^3N}{d^3P} =  \frac{1}{2\pi} \frac{d^2N}{p_t dp_t dy}
\left(1+2 \sum_n v_n\cos{(n \, \Phi)}\right)\ ,
\end{equation}
where $\Phi$ is the azimuth angle with respect to the true reaction plane
of the event.
In experiments anisotropic transverse flow manifests itself in the 
distribution of $\phi=\Phi+\Psi_R$, where $\phi$ is the 
measured azimuth for a track in detector coordinates, and 
$\Psi_R$ is the azimuth of the reaction plane in the event, which varies event
by event in the coordinate frame of the detector.
Using this definition the coefficients have a transparent meaning:
$$
v_n = \langle \cos(n (\phi - \Psi_R)) \rangle
$$
so that the directed flow is
$
v_1 = \langle p_x/p_t \rangle
$
and the elliptic flow is 
$
v_2 = \langle (p_x/p_t)^2 - (p_y/p_t)^2 \rangle
$ ,
where the average is taken over all emitted particles
(in a given rapidity, $y$, and transverse momentum $p_t$ bin),
in all events.
$\Psi_R$ can not be directly measured, and randomly takes any value
in $[0,2\pi]$ due to the random direction of the impact parameter vector
of the event. 

Anisotropic flow has been fully recognized as an important observable providing 
information on the early stages of heavy ion collisions \cite{Sorge,ZGYK,Ol92}. 
The development of flow is 
closely related to the pressure gradients, and thus, to the equation of state of the
nuclear matter formed in the collision \cite{RR97,Shu99}. 
Collective flow is also believed to be a promising signal to detect the creation of 
the quark-gluon plasma \cite{LaDi99,Brac00,BS02}. An increased 
attention to collective flow has also resulted in 
significant improvements in the techniques and methods of analysis and presentation 
of the experimental data. PHENIX, PHOBOS and STAR experiments at 
RHIC produced a wealth of information on the flow components 
\cite{STAR01,STAR02,STAR03,STAR04,PHO02,PHO03,PHE02,PHE03}.

\section{The Soft Point}

The most straightforward and first observed collective flow phenomenon
was the directed transverse flow. The projectile and target
in the collision have overlapped, the participant region is
compressed and heated up, which results in high pressure, $P$.
The spectator regions, which are not compressed and heated up so
much have contact with the hot central zone. The size of the
contact surface, $S$, is of the order of the cross-section of the
nuclei or somewhat less, depending on the impact parameter.
During the collision the spectators and participants were in contact
for a period of time $\tau$ and in this time the spectators
acquired a directed transverse momentum component of
$$
p_x \approx  P \times S \times \tau
$$
At a few hundred MeV/nucleon Beam energy the directed transverse
flow momentum was almost as large as the random average transverse
momentum and it was also close to the CM beam momentum.\cite{book}

The effect was dominant, it was used to estimate the compressibility
of nuclear matter, and the general expectation was that this is a good
tool to find the threshold of the phase transition of the
Quark Gluon Plasma. The reason is that the pressure decreases considerably
versus the energy density in a first order phase transition, when the 
phase balance and formation of the new phase prevents the pressure
from increasing in the mixed phase region. All estimates indicated that
this is a strong and observable decrease in the pressure, and
it has got the name the "soft point".

To some extent the expectations were fulfilled, the directed transverse
flow decreased as the beam energy was increased above 1 GeV and further.
This was obvious compared to the beam momentum, but the decrease was
also well observable compared to the average transverse momentum.

The reason is actually simple: although at high energies the pressure
should increase the target and projectile are becoming more and more Lorentz
contracted, and the overlap time also. So, the directed transverse momentum
the non-par\-ti\-ci\-pant matter could acquire was
$$
p_x \approx  P \times \frac{S}{\gamma} \times \frac{\tau}{\gamma} \ .
$$
Thus, at high energies the directed flow could not compete with the
trivial effect of reducing the flow by $\gamma^2$. This eliminated the 
interest in the directed flow, while the elliptic flow became a dominant
and strong effect.

\section{The Initial State}

The elliptic flow, which is observable 
in the coefficient of the 2nd harmonic, $v_2$, \cite{PV98} of 
the flow analysis, was not effected by the increasing beam energy
because it developed in the CM frame in the participant matter, due to
a special symmetry in the initial state and due to the large pressure
gradient in the direction of the reaction plane. The elliptic flow
took over the role of the directed flow, and became an important tool
in determining the basic reaction mechanism.  Only models, which
included a large collective pressure could reproduce the data, i.e.
mainly fluid dynamical models.  On the other hand, the elliptic flow
effect was so strong that most fluid dynamical models could fit the
data, even rather simple, one- and two- dimensional ones, so it was
not a very strong diagnostic tool.

A deeper insight into fluid dynamical calculations indicated that 
a similar effect, arising exclusively from participant matter, should
be possible to see also in the $v_1$ harmonic. This was first observed 
in 1999 \cite{LaDi99} based on earlier FD calculations, which included
a QGP EoS. The effect was named the 3rd flow component.

The effect was verified at the CERN SPS, but it was initially not
detected at RHIC. Now at the beginning of this year, first STAR then
two other collaborations have succeeded to measure the $v_1$ harmonic coefficient
of the collective flow at RHIC also. 
\cite{STAR04,Oldenburg} 
This indicated to us that one needs
extended theoretical studies, to map the sensitivity of this measurable.

As all fluid dynamical effects it depends on the initial state and on the
EoS. It is of special interest how this effect depends on the initial
state because this may shed light on the mechanisms of the formation
of QGP in heavy ion reactions.

The initial state models in
our recent works are based on a collective (or coherent)
Yang-Mills model, a versions of flux-tube models.  This approach \cite{GC86} 
was implemented in a Fire-Streak geometry streak by streak and it was
upgraded to satisfy energy, momentum and baryon charge conservations
exactly at given finite energies \cite{MCS01,MCSe02}. The effective 
string-tension was different for each streak, stretching in the
beam direction, so that central streaks with more color (and baryon) charges
at their two ends, had bigger string-tension and expanded less, than peripheral
streaks. The expansion of the streaks was {\it assumed} to last until the
expansion has stopped. Yo-yo motion, as known from the {\it Lund-model} was
not assumed. 

Our calculations show that a tilted initial state is formed, 
which leads to the 
creation of the third flow component \cite{LaDi99}, peaking at rapidities
$|y| \approx \pm 0.75$ \cite{MCSb01}. Recent, STAR $v_1$ data \cite{STAR04} 
indicate that our {\it assumption} that the
string expansion lasts until full stopping of each streak, may
also be too simple and local equilibration may be achieved earlier,
i.e. before the full uniform stopping of a streak. We did not 
explicitly calculate dissipative processes, some friction within and
among the expanding streaks is certainly present and experiments seem 
to indicate that this friction is stronger. 
The expanding strings are shown in Fig. 1

With time the streaks expand and move to the right and left on the
top and bottom respectively. This motion will be reflected in the
subsequent FD motion also so the tilted transverse expansion will be
observed at large polar angles, i.e. relatively low rapidities.
The position of the 3rd component flow peaks depends on the
(i) impact parameter, (ii) the effective time of the left/right
longitudinal motion, (iii) the string tension determining the 
lengths of the streaks, (iv) the thickness of the configuration which
determines the initial pressure gradient. The string tension varies
as a function of the distance of a given streak from the central beam
axis, because of the amount of the matter at the ends of the streaks.
This determined the amount of color charge, and so the string tension
is
$$
\sigma = A \left( \frac{E_{cm}}{m} \right)
 \frac{\sqrt{N_1 N_2}}{\delta x  \ \delta y}
$$
given in terms of the Baryon charges at the two ends of the streak, $N_1$ and $N_2$,
and the cross section of the string, $ \delta x\  \delta y$.

%----------------------FIGURE--------------------------------%

\begin{figure}[htb]
 \includegraphics[height=0.55\textheight]{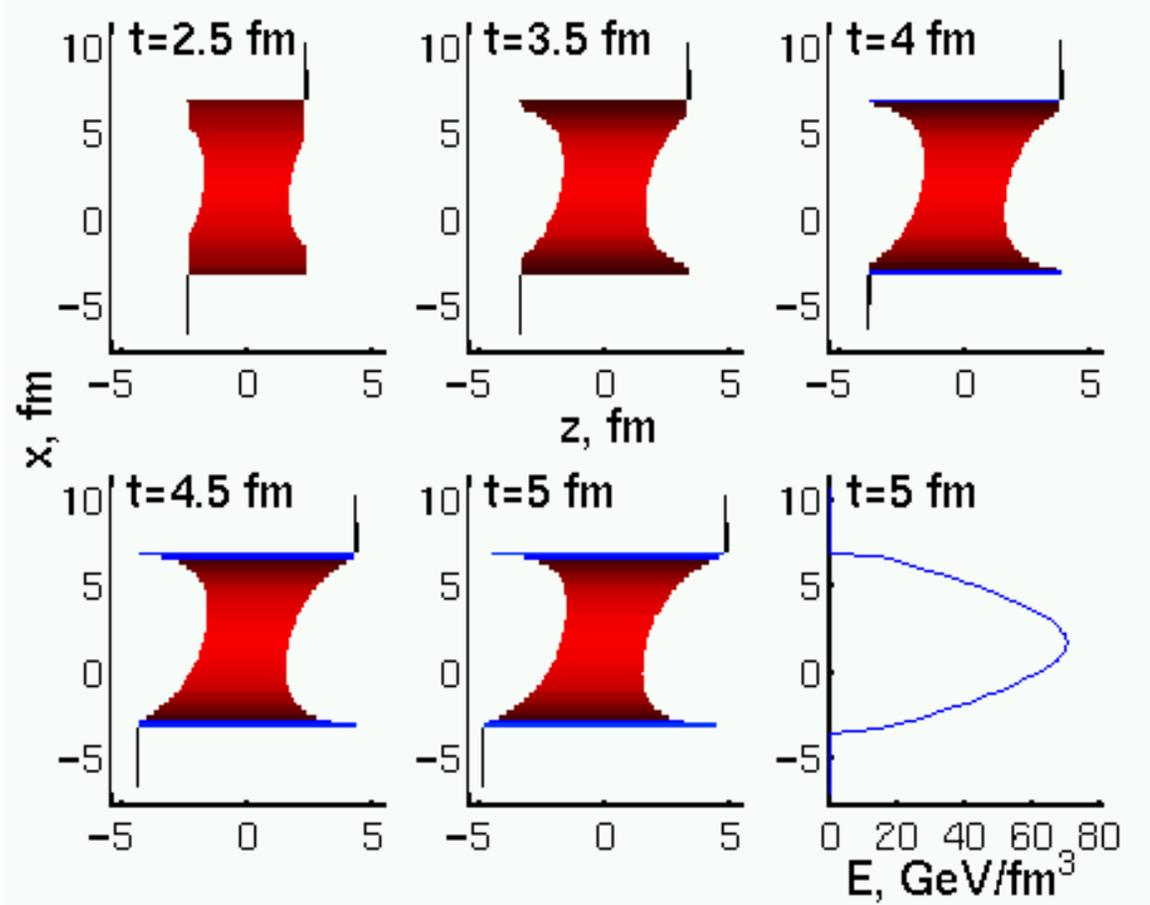}
\caption{
Au + Au collision at $E_{cm} = 65+65$ GeV/nucl, $b = 0.25 \times 2R_{Au}$,
and the parameter determining the string tension $A = 0.08$, 
$E = T_{00}$ is presented in the reaction plane as a function 
of $x$ and $z$ for different times in the laboratory
frame. The final shape of the QGP volume is a tilted disk $\approx 45^0$, 
and the direction of the fastest expansion will
deviate from both the beam axis and the usual transverse flow direction, 
and will generate the third flow component.
Note that the streaks are moving in the CM frame. From \cite{MCSe02}
}
\label{Cs-fig1}
\end{figure}

Although, there are several effects determining the angle of the 3rd flow 
component, some of these can be traced down by other measurements also:
the impact parameter by the multiplicity, the effective expansion time 
and size by two particle correlation measurements, etc. Thus, there is a
reasonable possibility that the $v_1$ measurements will lead to a more detailed 
insight to the details of the QGP formation in heavy ion reactions. 

The subsequent fluid dynamical calculations show the development of the
3rd flow component from these type of initial states. These FD model calculations
have to be supplemented with a Freeze Out model. There is a very essential
development in this field also, and we can have different levels of approach
from simplified freeze out descriptions across an FO hypersurface using
an improved Cooper-Fry description, or a full kinetic description, originating
from a modified Boltzmann Transport Equation approach. This final step
is complicated further by the fact that the data indicate a rapid and simultaneous 
hadronization and freeze out. 

Nevertheless, we expect that the $v_1$ flow data will be sufficiently detailed
and accurate soon, and the 3rd flow component will be an effect which is
sufficiently strong that the complex final effects at the freeze out
will allow a successful analysis of the matter properties via these flow
measurements.

\section{Analysis of flow components in experiments}\label{experiment}  

There are several techniques, which have been used to calculate flow components. One of the
most important, however, far not trivial task is the identification of the reaction plane. 
The Danielewicz-Odyniecz method \cite{DO85} constructed an estimated event plane
$\Psi_{\{EP\}}$,
using the the momentum vectors of all detected particles, introducing a
rapidity dependent weighting (where the sign of the CM rapidity
was crucial), and by eliminating self correlations. This weighting
in principle should also be used in the Fourier expansion method
\cite{PV98}, especially for odd harmonics.
When the Fourier expansion method is used, each measurable harmonic can
yield an independent estimated $\Psi_n$, event plane of the $n$th harmonic. These estimated
 reaction planes may differ from one-another, which is clearly incorrect, as there is only
 one reaction plane in one event.
In some of the recent techniques the reaction plane is not evaluated explicitly, because 
the Fourier coefficients, $v_1$, $v_2$, $v_3$, ... can be obtained without. We will show 
that these methods may lead to problems as well.

\subsection{Techniques for analyzing $v_n$}\label{techniques}

In this section we will briefly discuss three of the experimental methods of flow analysis.

The flow coefficients can be obtained by the {\it pairwise correlation} \cite{WANG91} of 
all particles without referring to the reaction plane. This two-particle correlation 
method produces the squares of the coefficients: 
$$
v_n^{2}=\langle\cos{[n(\phi_i-\phi_j)]}\rangle_{i \neq j} \ .
$$ 
The method has the advantage that the reaction plane does not need to be determined 
or estimated by an event plane.

In the {\it event plane method} \cite{PV98} one investigates the correlation of particles 
with an event plane, which is an estimation to the real reaction plane. The flow components 
are given by:
$$
v_n^{obs}=\langle\cos{[n(\phi_i-\Psi_n)]}\rangle \ ,
$$    
where $\Psi_n$ is the observed event plane of order $n$. The observed event plane is not the 
true reaction plane, therefore, the observed coefficients, $v_n^{obs}$, have to be corrected 
by dividing by the resolution of the event plane. The resolution is estimated by measuring 
the correlations of the {\it event planes of sub-events}. For details consult \cite{PV98} 
or \cite{PHO02}. 
 
The sub-event method is used originally in the Danielevicz-Odyniecz
method \cite{DO85} to determine the accuracy of the estimated reaction
plane from the data. Using it in the Fourier expansion method, there are several ways to 
choose sub-events. Most trivially one can divide each event randomly into
two sub-events. The partition using two (pseudo) rapidity
regions (better separated by $\Delta y > 0.1$) could greatly
suppress the contribution from quantum statistics effects and
Coulomb (final state) interactions \cite{STAR02}, as well as could also
be used to correctly identify the first harmonic reaction plane, $\Psi_1$.

Recently, a {\it multiparticle correlation} or {\it cumulant method} \cite{BORG01,BORG02} is 
widely used. This method has larger statistical errors than the two-particle analysis. 
The most recent STAR data on the directed flow \cite{STAR04} were 
calculated with three-particle cumulants combined with the {\it event plane method}.

\subsection{Recent detection techniques}\label{problems}

In experimental techniques using the Fourier expansion method, each measurable harmonic can
yield an independent, estimated $\Psi_n$, via the event flow vector
$Q_n$ with the definition 
\cite{STAR02}:
$$
Q_n\ \cos(n \Psi_n) = \Sigma_i \cos(n \phi_i), \ \ \
Q_n\ \sin(n \Psi_n) = \Sigma_i \sin(n \phi_i)\ ,
$$
where the sums extend over all particles in a given event.

First of all these estimated reaction planes may be different 
from one-another, furthermore as the summation goes over the 
whole acceptance of the detector, which is symmetric in rapidity,
$y$ \cite{STAR02} without weighting by rapidity, the {\it first harmonic}, $v_n$, 
is eliminated by construction, because it involves a Forward/Backward
azimuthal antisymmetry, and so, the Forward and Backward contributions
cancel each other in the above definition.

One might argue that such a Forward/Backward azimuthal anti-correlation is
a consequence of momentum conservation, thus, it is a non-flow correlation, 
but fluid dynamics is nothing else
then the collective form of energy and momentum conservation! 

More precisely, in the infinite particle number limit, fluid dynamics really leads to 
a single particle momentum distribution after integrating the contributions of all fluid 
elements. This is a consequence of the assumption of local local equilibrium, a fundamental 
assumption in fluid dynamics, and of the assumption of molecular chaos. When we consider 
{\it finite multiplicities} and smaller samples, correlations may arise from global momentum
 conservation. To subtract these correlations as non-flow effects is questionable. 
Furthermore, fluid dynamics must be supplemented by some {\it Freeze Out} (FO) prescription
 to obtain measurables.

Due to the Freeze Out process, the  local thermal equilibrium ceases to exist, the post FO 
distribution must be an out of equilibrium, non-thermal distribution. In the FO process the 
assumption of molecular chaos does not hold, so the FO process leads to correlations. 
There is a third effect, inherent in fluid dynamical descriptions when {\it sudden and 
rapid hadronization} coincides with freeze out. This can be described in a non-thermal 
string fragmentation, coalescence or recombination picture, which lead to correlations also. 
The above mentioned three effects fundamentally influence the measured flow patterns, and 
the measured Fourier harmonics, so it is highly questionable if these should be excluded 
from the determination of the reaction plane, as it is done in the cumulant expansion method.

For the higher {\it odd harmonics}, the determination of reaction plane  can 
be similarly problematic, as these appear dominantly in asymmetric,
$A+B$, collisions, and then the beam direction is crucial, and the
method eliminates this important physical information. Thus, the elimination
of the information provided by the longitudinal motion of emitted
particles and the longitudinal symmetries and asymmetries severely
impair the analysis of the collective flow.

In case of {\it even harmonics} exclusively, the weighting does not seem to be too
important and one may conclude (wrongly) that it can be omitted without severe
consequences. For even harmonics, however, there is a symmetry for 
positive and negative $x$-values, and so, the Projectile/Target directions
cannot be identified, i.e. only the reaction plane is identified but not
the direction of the impact parameter vector ${\mathbf b}$. If this estimated
reaction plane is used for the evaluation of the coefficients, $v_n$, the
Target and Projectile directions may appear randomized in the sample.
If weighting is used, but it is Forward/Backward symmetric, e.g. $w_i (y) = v_2(y)$
\cite{STAR02},
this does not solve the problem discussed here.
As a consequence Forward/Backward azimuthal asymmetries, will be decreased 
or eliminated
by this misidentification, and these will contribute
and enhance the measured coefficients of even harmonics, e.g. the 
elliptic flow.

The above mentioned situation may occur even if the reaction plane 
determination is implicit and not discussed at all.
Due to complicated experimental setups, the event plane determination
varies to a large extent, and different methods can even be mixed with
each other. In these cases it is very difficult to judge the
accuracy and precision of event plane determination. An example is, the 
evaluation of $v_1$ by the STAR collaboration, where 
two different methods determining the reaction plane yield
different coefficients \cite{STAR04,Oldenburg} $v_1 = (-0.25 \pm 0.27(stat)) \% y$ vs.
$v_1 = (-0.5 \pm 0.5(stat))\% y$, which agree within error and are even consistent with
zero within error.

\section{Calculation of flow components}\label{calculation}

We have calculated the directed and elliptic flow from a tilted, ellipsoidally expanding 
particle emitting source. Our tool is a simple, blast wave type hydrodynamic model.   
The tilt angle, $\Theta$, represents the rotation of the major (longitudinal) direction 
of expansion from the direction of the beam. In the presented calculations 
$\Theta = \pm 5.7^{\circ}$.
We have divided our fireball into cubic cells by a grid in x, y, z coordinates, as it is 
done in most hydrodynamic models. The aim of the introduced discretization was to produce 
similar output as other models have, which makes it possible to change the presently used 
simple blast wave model to more sophisticated ones without further changes in the next steps 
of the calculation.

Also, the FO layer is discretized on this grid. Due to this discretization, the 
``fluid-cells'' do not match the spherical layer exactly, the volume of the cells, 
and so all conserved quantities have some discretization error. This depends 
on the choice of radius, layer thickness, cell size and the way which cells 
are selected to be in the layer. However, one can vary these parameters, 
to achieve a small relative error in the normalization. The present example 
yields a relative error below $1 \% $ which is already much smaller than the 
statistical errors of the experimental techniques.

\subsection{Theoretical background}\label{theory}

The contribution of a fluid cell to the final baryon Phase-Space (PS)
distribution is:
\begin{equation}
\frac{dN_{c}}{d^3p} = \gamma \, V_{c} \,  \:
\frac{p^{\mu} \, d\sigma_{\mu}}{p^0} \ \ f_{F.O.}(x,\, p) \ ,
\label{def1}
\end{equation}
where $\gamma \, V_{c}$ is the proper volume of one fluid 
cell, $f_{F.O.}(x,\, p)$ is the freeze out distribution and $d\sigma_{\mu}$ is the normal 
of the FO surface.
Using the relations
$p = (p^0, p_\parallel, {\mathbf p_\perp})$,
$p_\parallel = p^0 \ dy$ and
$p_t = | {\mathbf p_\perp} | $, we can get the azimuthal distribution per unit
rapidity as
$$
\frac{dN_{c}}{dy\ d\phi} =
\int \frac{d^3 p}{dy\ d\phi}\ \frac{dN_{c}}{d^3p}   =
\gamma \, V_{c} \,
\int \frac{d^3 p}{dy\ d\phi}\ \frac{p^{\mu}\, d\sigma_{\mu}}{p^0} \
f_{F.O.}(x,\, p) =
$$
\begin{equation}
= \gamma\, V_{c}\, \int dp_t\ p_t\ \ (p^{\mu}\, d\sigma_{\mu})\ f_{F.O.}(x, p)
\equiv \ \int dp_t\ p_t\ G_c (p_t, \phi_{CM}, y)\ .
\label{ptint}
\end{equation}

When we evaluate the azimuthal asymmetry this is done with respect to the
reaction plane. The $\phi_{CM}$ is the azimuth-angle of particles in the
C.M. frame where these are measured. Then, the coefficients of
the different harmonics, $v_1$, $v_2$, etc. can be evaluated via additional
numerical integrations over $\phi_{CM}$ azimuth angle. 

Thus,
\begin{equation} 
v_n(y) = 
\frac{
\displaystyle{{ \sum_{c} \int  \cos{(n\phi_{CM})} \, 
\gamma\, V_{c}\,(p^{\mu}\, d\sigma_{\mu})\ f_{F.O.}(x, p) d^2p_t }}
}{
\displaystyle{\sum_{c}\ dN_c / dy}} \ .
\label{v_n}
\end{equation}
For the FO surface we may assume that the local momentum distribution  is a J\"uttner distribution: 
$$
f_{F.O.}(x,\, p) = f^{J\ddot{u}ttner}(p) \equiv \frac{g_{n}}{(2 \pi
  \hbar)^3} exp \left( \frac{\mu -p^{\mu}u_{\mu}}{T} \right) \ .
$$
In this case Eq. (\ref{v_n}) takes the form
\begin{eqnarray} 
v_n(y) =
\frac{
\displaystyle{ \sum_{c} \, K \, \gamma V_c\  
\int dp_t\ p_t\ d\phi_{CM} \,  \cos{(n \, \phi_{CM})}
g_{c}(p_t, \phi_{CM}, y) }
}{
\displaystyle{\sum_{c}\ dN_c / dy}} \nonumber 
\end{eqnarray}
where we have introduced the following notations:
\begin{eqnarray}
g_{c}(p_t, \phi_{CM}, y) \equiv \left\lbrack H \,\sqrt{m^2 + p_{\bot }^2}  - 
 \, {\rm\bf {p}_{\bot}} \, \gamma_{\sigma} \: {{\rm\bf d\sigma}_{\bot}} \: \right\rbrack \cdot
  e^{- h \,\sqrt{m^2 + p_{\bot }^2}  + \, {\rm\bf {p}_{\bot}} \, {\rm\bf g} } \nonumber \ , \\
%\end{eqnarray}
%
%\begin{eqnarray}
H \equiv \gamma_{\sigma} (\cosh y - d\sigma_{\parallel} \, \sinh y )\ , \quad 
h \equiv \gamma (\cosh y - v_{\parallel} \, \sinh y )/ T \ ,\quad 
 {\rm\bf g}  \equiv \gamma \: {\rm\bf v_{\bot}}/T  \nonumber \ ,\\
K \equiv (g_n \cdot e^{ \, \mu / \, T})/(2 \pi \hbar)^3 = (g_n
  \cdot n)/(4 \pi \, m^2 \, T \, K_2(m/T)) \nonumber \ .
\end{eqnarray}

To calculate the $v_n$ harmonics one has to perform double integrals. 
The calculation of the numerator can only be done  numerically. However, the denominator, 
which is actually the rapidity distribution of particles, $dN/dy$, has an analytical 
solution. The solution for the case of $d\sigma^{\mu} = u^{\mu}$ was derived and shown 
in \cite{book} in Eq. (7.6). For the more realistic, general case, when $d\sigma^{\mu} \not=
u^{\mu}$ such analytical result, according to our knowledge, has not yet been shown in 
the literature. We have, however, found such solution in this latter case as well,
\cite{NyH04} 
and derived a relatively simple formula to calculate the rapidity distribution, i.e. the 
denominator of the $v_n$ flow components:
\begin{eqnarray}
\frac{dN_{c}}{dy} &= & 2 \pi \, K \, \gamma  V_c \,\gamma\prime^3 \, \frac{H}{h} m^2 \, \left( 1-
\frac{g \, G}{h \, H}\right)
\left\lbrack  
\frac{2 \, \gamma\prime^2}{h^2 \, m^2} + \frac{2 \,
  \gamma\prime}{h \, m} + 1  \right\rbrack 
e^{-\frac{h}{\gamma\prime}  m} \ .
\label{dndy}
\end{eqnarray}
>From Eq. (\ref{dndy}) the final rapidity distribution can be calculated by summing over 
all fluid cells, i.e. $dN/dy = \sum_c dN_{c}/dy$. The new formula makes further calculations 
faster, because we can reduce the number of time consuming numerical integrations.

\subsection{Results for directed and elliptic flow}\label{results}

Our primary aim is to show why the identification of the reaction plane is so important, and how 
the odd harmonics may be eliminated by construction using the cumulant method without proper 
weighting. As we have mentioned in Sec. \ref{problems}, the acceptance of the detector is symmetric in 
rapidity \cite{STAR02}. Odd harmonics involve a Forward/Backward azimuthal antisymmetry, 
therefore, without weighting by rapidity taking into account the sign of it, i.e. whether the 
detected particle came from the target or projectile side, the Forward and Backward contributions
may cancel each other. To show this, we have calculated flow components from two tilted ellipsoidally 
expanding sources, which differ from one-another only in the sign of the tilt angle, i.e. 
we have changed the 
projectile and target side. Then we calculated the average of both the directed and elliptic 
flow components coming from the two opposite tilted sources, which simulates the situation when 
only the reaction plane is identified, but not the impact parameter, as the projectile and 
target directions are not known.  
 
\begin{figure}[htb]
\vspace*{-1.1cm}
 \includegraphics[height=0.55\textheight]{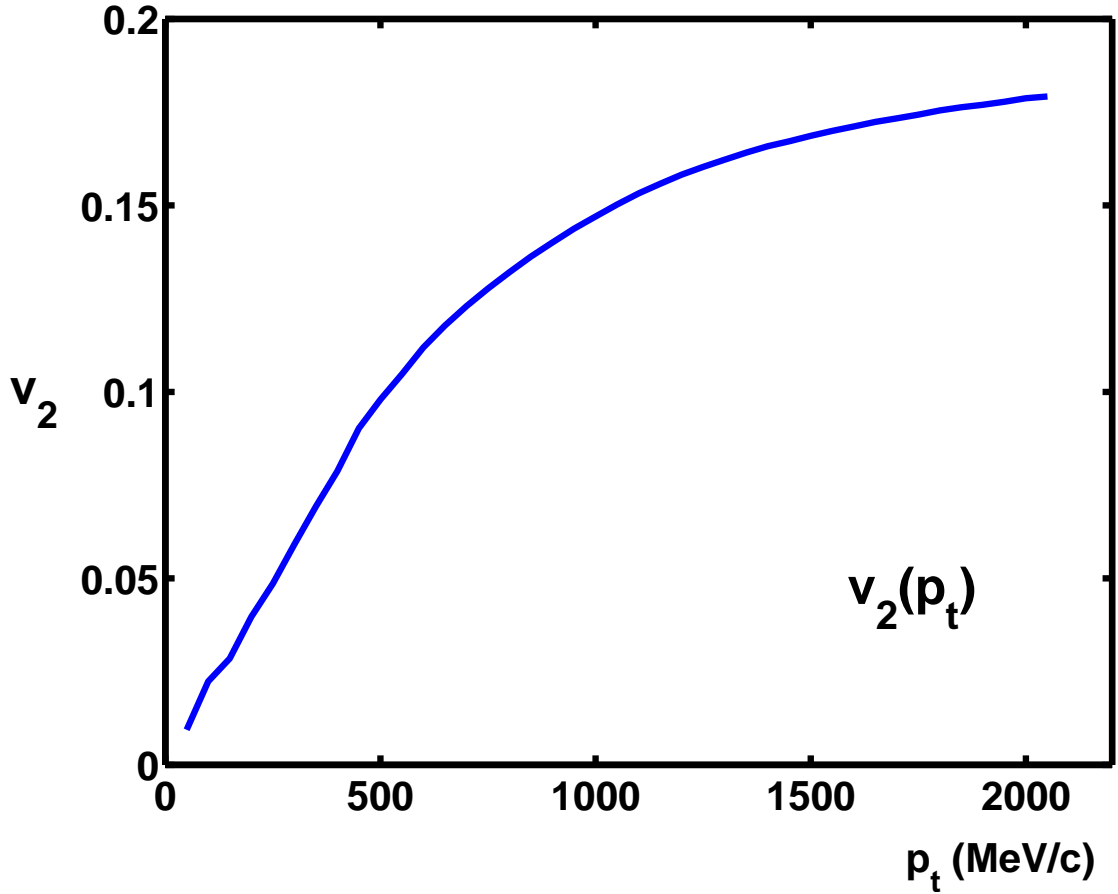}
\vspace*{-2.1cm}
\caption{Elliptic flow of nucleons, $v_2$, as a function of transverse momentum, $p_t$, 
at mid-rapidity, $y=0$, from \cite{NyH04}. The the velocity is normal to the tilted ellipsoid,
with axes $a, b, c = 6, 8, 10 fm$. The sudden FO happens with $T=165 MeV$ and $n=0.5 fm^{-3}$. 
The result is in good agreement with the experimental 
data presented in \cite{STAR04}.}
\label{fig1}
\end{figure}

Figure \ref{fig1} shows the elliptic flow, $v_2$, result as a function of the transverse 
momentum, $p_t$. These figure is less relevant for our primary aim, but demonstrates that even 
a simple blast wave model can reproduce some of the main characteristics of the observed data. 
We have plotted a less wide  $p_t$ region, because it was shown earlier \cite{STAR03} that 
elliptic flow at RHIC can be described by hydrodynamical models for $p_t$ up to $2 GeV/c$. 
$v_2$ rises almost linearly up to $p_t = 1 GeV/c$, then deviates from a linaer rise and starts to saturate.

%----------------------FIGURE--------------------------------%

\begin{figure}[htb]
 \includegraphics[height=0.55\textheight]{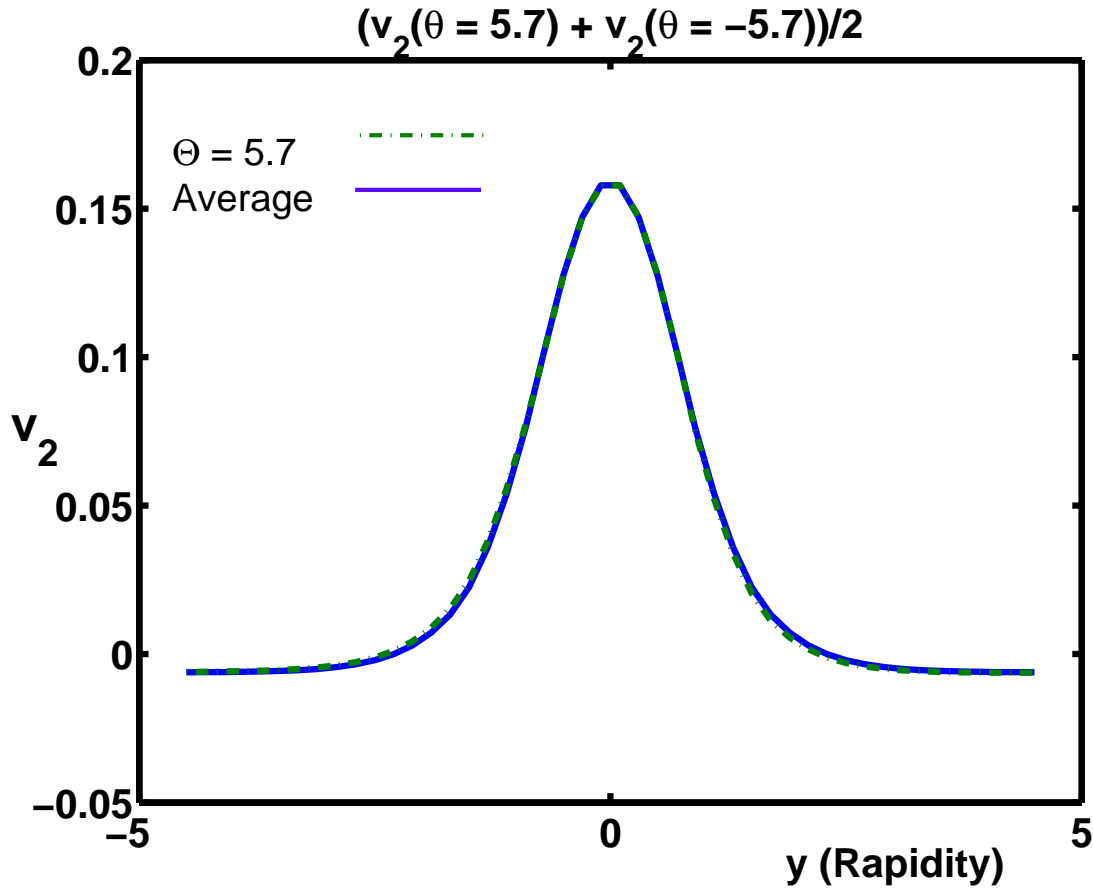}
\caption{Elliptic flow, $v_2$, as a function of rapidity, $y$. The distribution is somewhat 
narrow. This problem can be avoided by better choice of parameters. One can see that 
the $v_2$ result is not sensitive too the change of projectile and target side, 
the two curves are almost identical.}
\label{fig2}
\end{figure}

Figure \ref{fig2} demonstrates that identification of the Projectile/Target plays a less 
important role in the determination of the second harmonic, such as higher order even harmonics.
For even harmonics there is a symmetry for positive and negative $x$-values, thus the role of 
introducing weights with opposite signs for positve and negative rapidities in the event plane 
or cumulant method is not transparent. This may lead to the wrong conclusion that weighting is 
not important. 
In Figure \ref{fig2} we have plotted the elliptic flow , $v_2$, as a function of rapidity, 
$y$. The distribution is too narrow compared to the expected one and to experimental data. 
However, it is possible to improve the recent result by using more suitable set of parameters. 
The dashed-dotted curve refers to the case when the tilt angle is positive, 
$\Theta = 5.7^{\circ}$, while the continuos line represents the result from 
averaging $v_2(\Theta = 5.7^{\circ})$ and $v_2(\Theta = -5.7^{\circ})$, i.e. 
elliptic flows from two oppositely tilted sources.  The two results are nearly identical, therefore 
one can hardly see that there are indeed two curves in the figure.

Let us perform a similar investigation for the directed flow, or first harmonic, $v_1$. We 
calculate $v_1$ as a function of rapidity, $y$, in the case of two oppositely tilted ellipsoidal 
sources. The dashed-dotted curve refers to the case with $\Theta = 5.7^{\circ}$.

%----------------------FIGURE--------------------------------%

One can see, 
that $v_1(y)$ is definitely not constant zero and the so called ``wiggle'', which is well known 
from earlier experiments with lower energies, appears. The situation changes dramatically when 
we construct the averaged $v_1$, which demonstrates what happens when we partly reverse the 
projectile and target side, i.e. we use the event plane or cumulant method without proper 
weighting. As the continuos curve shows, in this case $v_1$ is in principle set to zero, 
$v_1(y) \approx 0$. As we have mentioned at the beginning of Section 
\ref{calculation}, we have introduced a discretization of our freeze out layer. This discretization 
leads to the inaccuracy in our calculations. 
In other calculations with bigger tilt angles, $|\Theta| > 10^{\circ}$ 
such ``waves'' were not seen. However, theoretical considerations do not support  
that the source can be so much tilted at RHIC energies, therefore, we have chosen smaller 
angles in our presentation.

\section{Conclusions}\label{concl}

We have studied the importance of the initial state in the development
of cllective flow in heavy ion collisions. The directed transverse
flow weakens at increasingly ultra-relativistic energies, so that its
recovery after the soft point cannot be expected or easily demonstrated.

%----------------------FIGURE--------------------------------%

\begin{figure}[htb]
 \includegraphics[height=0.55\textheight]{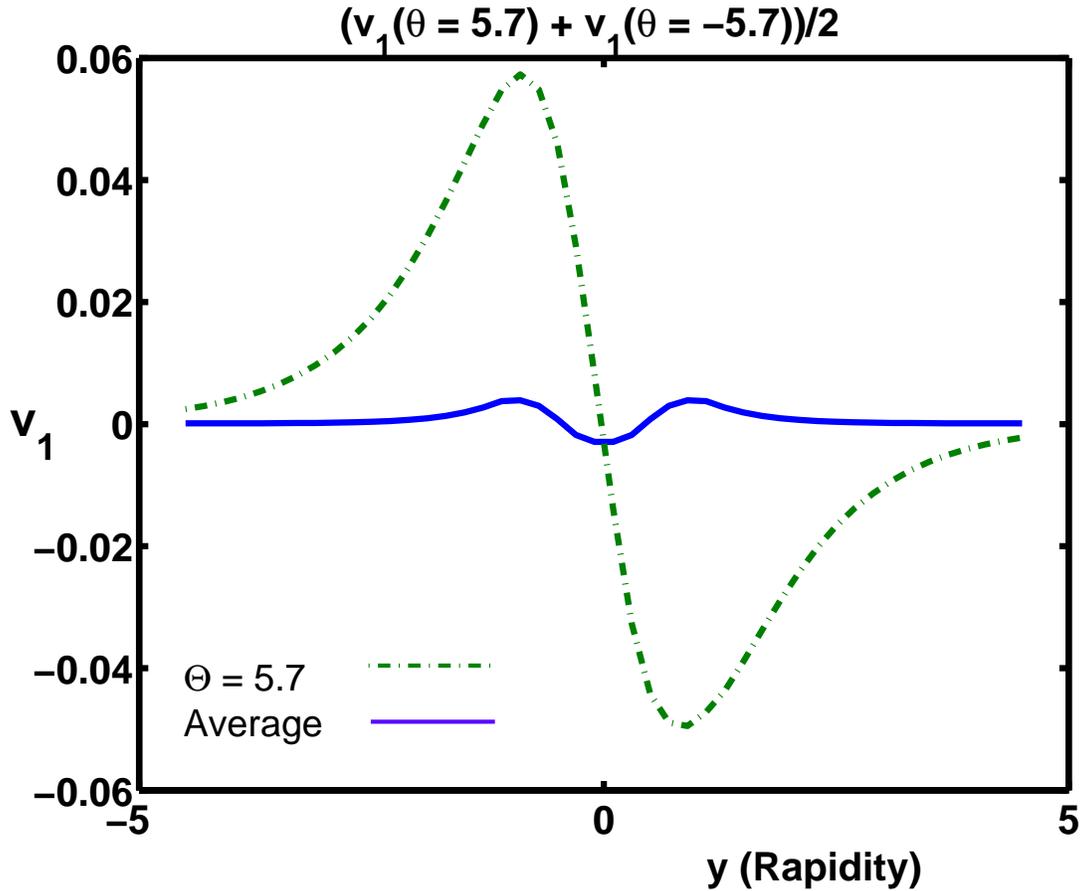}
\caption{Directed flow, $v_1$, as a function of rapidity. 
The dashed-dotted line refers to $v_1(\Theta = 5.7^{\circ})$, 
while the continous line is the result from averaging the 
directed flows calculated with $\Theta = 5.7^{\circ}$ and $- 5.7^{\circ}$.}
\label{fig3}
\end{figure}

On the other hand QGP formation leads to an elongated ellipsoidal initial
state which is clearly shown by the increasing elliptic flow, $v_2$. The longest
axis of this ellipsoid shape, however, must not and cannot be exactly
parallel to the beam axis, at finite impact parameters, therefore it leads to
a forward-backward azimuthal asymmetry, the 3rd flow component.  Careful
experimental analysis can identify this asymmetry in the $v_1$ flow harminics.

The analysis of this asymmetry provides extremely valuable information
on the length and tilt of the initial state of the QGP formed initially
in heavy ion collisions by the time local equilibration and thermalization
are reached. 

To study the sensitivity and detectability of the asymmetry arising from 
this flow we initiated calculations of lower flow harmonics. 
So far we have not calculated higher harmonics, but those calculations are 
starightforward using our model and Eq. (\ref{v_n}).  
We found that model results for the first 
even harmonic, $v_2$, are in good agreement with experimental data. 

However, the first odd harmonics, or directed flow, $v_1$, can be misinterpreted in some
of the experimental techniques, and the same may be true for the higher, odd 
harmonics, which have not been published yet. We have 
pointed out several critical points in the recently used experimental 
methods for calculation of flow componets, which may lead to problems in the flow 
analysis. These arise from the insufficiently accuracy of the identification of the 
reaction plane.

Therefore, further improvements of both the experimental techniques and theoretical 
models are needed. Especially, the reaction plane should be 
determined more accurately.  Further work should include the 
study  of energy dependence of flow components for different hadronic species, which 
could give information on pressure and pressure gradients in the nuclear matter 
created in the collisison. 

In conclusion we can state that hydrodynamic modeling
of heavy ion reactions is alive and is better than ever.
Clear hydrodynamic effects are seen everywhere, and from
early on.

This indicates we are approaching a regime where collective
matter type of behavior is dominant. We hope to gain
more and more detailed information on QGP and its dynamical
properties. Continued hard work is needed to exploit all
possibilities, and the task of theoretical modeling
and analysis is vital in future progress of the field.

\begin{theacknowledgments}

The authors thank the hospitality of the Frankfurt Institute of Advanced
Studies and the Institute for Theoretical Physics of the University of 
Frankfurt, the Gesellschaft f\"ur Schwerionenforschung, and the University
of Giessen where parts of this work were done.
L.P. Cs., thanks the Alexander von Humboldt Foundation
for extended support in continuation of his earlier Research Award.
\'A. Ny. wishes to thank Boris Tom\'a\u{s}ik and Halvor M. Nilsen 
for numerous very helpful discussions. 
This work was supported by the Norwegian Research Council.
 
\end{theacknowledgments}

%%%%%%%%%%%%%%%%%%%%%%%%%%%%%%%%%%%%%%%%%%%%%%%%
%% You may have to change the BibTeX style below, depending on your
%% setup or preferences.
%%
%% If the bibliography is produced without BibTeX comment out the
%% following lines and see the aipguide.pdf for further information.
%%
%% For The AIP proceedings layouts use either
%%%%%%%%%%%%%%%%%%%%%%%%%%%%%%%%%%%%%%%%%%%%

\bibliographystyle{aipproc}   % if natbib is available
%\bibliographystyle{aipprocl} % if natbib is missing

%%%%%%%%%%%%%%%%%%%%%%%%%%%%%%%%%%%%%%%%%%%
%% You probably want to use your own bibtex database here
%%%%%%%%%%%%%%%%%%%%%%%%%%%%%%%%%%%%%%%%%%%
\bibliography{Csernai}
%\bibliography{sample}

\hyphenation{Post-Script Sprin-ger}

%%%%%%%%%%%%%%%%%%%%%%%%%%%%%%%%%%%%%%%%%%%
%% Just a reminder that you may have to run bibtex
%% All of it up to \end{document} can be removed
%% if you don't like the warning.
%%%%%%%%%%%%%%%%%%%%%%%%%%%%%%%%%%%%%%%%%%%
\IfFileExists{\jobname.bbl}{}
 {\typeout{}
  \typeout{******************************************}
  \typeout{** Please run "bibtex \jobname" to optain}
  \typeout{** the bibliography and then re-run LaTeX}
  \typeout{** twice to fix the references!}
  \typeout{******************************************}
  \typeout{}
 }

\end{document}